\documentstyle[12pt]{article}

\newcommand{\be}{\begin{equation}}
\newcommand{\ee}{\end{equation}}
\begin{document}
\begin{titlepage}
\begin{flushright}
CERN-TH/95-138\\
UWThPh-19-1995\\
hep-th/9505175
\end{flushright}
\begin{center}
{\Large \bf Towards Finite Quantum Field Theory}\\
{\Large \bf in Non-commutative Geometry}\\[20pt]
H. Grosse\footnote{Part of Project No. P8916-PHY
of the `Fonds zur F\"orderung der wissenschaftlichen Forschung
in \"Osterreich'.}  \\
{\small Institut for Theoretical Physics, University of Vienna,\\
Boltzmanngasse 5, A-1090 Vienna, Austria \\}
C. Klim\v{c}\'{\i}k\footnote{Partially supported by the grant GA\v CR
2178} \\
{\small Theory Division CERN,
CH-1211 Geneva 23, Switzerland \\}
P. Pre\v{s}najder \\
{\small Department of Theoretical Physics, Comenius University \\
Mlynsk\'{a} dolina, SK-84215 Bratislava, Slovakia\\}
{\bf Abstract} \\
\end{center}

We describe the self-interacting scalar field on the truncated
sphere and we perform the quantization using the functional (path)
integral approach. The theory posseses a full symmetry with respect to
the isometries of the sphere. We explicitely show that the model is
finite and the UV-regularization automatically takes place.

\noindent CERN-TH/95-138 \\
May 1995
\end{titlepage}

\section{Introduction}
The basic ideas of the non-commutative geometry were developed in [1, 2],
and in the form of the matrix geometry in [3, 4]. The applications to
physical models were presented in [2, 5], where the non-commutativity
was in some sense minimal: the Minkowski space was not extended
by some standard Kaluza-Klein manifold describing internal degrees of freedom
but just by two non-commutative points. This led to a new insight on the
$SU(2)_L \bigotimes U(1)_R$ symmetry of the standard model of electro-week
interactions. The model was further extended in [6] inserting the
Minkowski space by pseudo-Riemannian manifold, and thus including the
gravity.
Such models, of course, do not lead to UV-regularization, since they do not
introduce any space-time short-distance behaviour.

To achieve the UV-regularization one should introduce the non-commuta-
\newline
tivity into the genuin space-time manifold in the relativistic case, or into
the space manifold in the Euclidean version. One of the simplest locally
Euclidean manifolds is the sphere $S^2$. Its non-commutative (fuzzy) analog
was described by [7] in the framework of the matrix geometry. More general
construction of some non-commutative homogenous spaces was described in [8]
using coherent states technique.

The first attempt to construct fields on a truncated sphere were presented in
[9] within the matrix formulation. Using more general approach the classical
spinor field on truncated $S^2$ was investigated in detail in [10-11].

In this article article we shall investigate the quantum scalar field $\Phi$
on the truncated $S^2$. We shall explicitely demonstrate that the
UV-regularization automatically appears within the context of the
non-commutative geometry. We shall introduce only necessary notion of the
non-commutative geometry we need in our approach. In Sec. 2 we define the
non-commutative sphere and the derivation and integration on it. In
Sec. 3 we introduce the scalar self-interacting field $\Phi$ on the truncated
sphere and the field action. Further, using Feynman (path) integrals we
perform the quantization of the model in question.
 Last Sec. 4 contains a brief discussion
and concluding remarks.

\section{Non-commutative truncated sphere}
A) The infinite dimensional algebra ${\cal A}_{\infty}$ of polynomials
generated by $x = (x_1, x_2, x_3) \in {\bf R^3}$ with the defining relations
\be
[x_i, x_j] = 0, \ \ \ \sum_{i=1}^3 x_i^2 = \rho^2
\ee
contains all informations about the standard unit sphere $S^2$ embedded
in ${\bf R^3}$. In terms of spherical angles $\theta$ and $\varphi$ one has
\be
x_{\pm} = x_1 \pm i x_2  = \rho e^{\pm i\varphi} \sin \theta \ ,\ \
x_3 = \rho \cos \theta \ .
\ee

As a non-commutative analogue of ${\cal A}_{\infty}$ we take the algebra
${\cal A}_N$ generated by $\hat{x} = (\hat{x}_1, \hat{x}_2, \hat{x}_3)$ with
the defining relations
\be
[\hat{x}_i, \hat{x}_j ] = i \lambda \varepsilon_{ijk} \hat{x}_k \ ,\ \
\sum_{i=1}^3 \hat{x}_i^2 = \rho^2 \ .
\ee
The real parameter $\lambda > 0$ characterizes the non-commutativity
(later on
it will be related to $N$). In terms of
$\hat{X}_i = \frac{1}{\lambda} \hat{x}_i \ , i=1,2,3$, eqs. (3) are changed
to
\be
[\hat{X}_i, \hat{X}_j ] = i \varepsilon_{ijk} \hat{X}_k \ ,\ \
\sum_{i=1}^3 \hat{X}_i^2 = \rho^2 {\lambda}^{-2} \ ,
\ee
or putting $X_{\pm} = X_1 \pm i X_2$ we obtain
\be
[\hat{X}_3, \hat{X}_{\pm} ] = \hat{X}_{\pm} \ ,\ \
[\hat{X}_+, \hat{X}_- ] = 2 \hat{X}_3 \ ,
\ee
and
\be
C = \hat{X}_3^2 + \frac{1}{2} (\hat{X}_+ \hat{X}_- + \hat{X}_- \hat{X}_+ )
= \rho^2 {\lambda}^{-2} \ .
\ee

We shall realize eqs. (4), or equivalently eqs. (5) and (6), as relations in
some suitable irreuducible unitary representations of the $SU(2)$ group. It
is
useful to perform this construction using Wigner-Jordan realization of the
generators $\hat{X}_i \ , i=1,2,3$, in terms of two pairs of annihilation and
creation operators $A_{\alpha} , A^*_{\alpha} , \alpha = 1,2$, satisfying
\be
[A_{\alpha} , A_{\beta} ] = [A^*_{\alpha} , A^*_{\beta} ] = 0 \ ,\ \
[A_{\alpha} , A^*_{\beta} ] = \delta_{\alpha ,\beta} \ ,
\ee
and acting in the Fock space $\cal{F}$ spanned by the normalized vectors
\be
|n_1 ,n_2 \rangle =
\frac{1}{\sqrt{n_1 ! n_2 !}} (A^*_1)^{n_1} (A^*_2)^{n_2} |0 \rangle \ ,
\ee
where $|0 \rangle$ is the vacuum defined by $A_1 |0 \rangle = A_2 |0 \rangle
= 0$. The operators $\hat{X}_{\pm}$, and  $\hat{X}_3$ take the form
\be
\hat{X}_+ = 2A^*_1 A_2 \ ,\ \hat{X}_- = 2A^*_2 A_1 \ ,\
\hat{X}_3 = \frac{1}{2} (N_1 - N_2 ) \ ,
\ee
where $N_{\alpha} = A^*_{\alpha} A_{\alpha} \ , \alpha = 1,2$. Restricting to
the $(N+1)$-dimensional subspace
\be
{\cal{F}}_N = \{ |n_1 , n_2  \rangle \ \in \cal{F} \} \ ,
\ee
we obtain for any given $N\ =\ 0,1,2,...\ \ $, the irreducible unitary
representation in which the Casimir operator (6) has the value
\be
C = \frac{N}{2} \left( \frac{N}{2} +1\right) \ ,
\ee
i.e. the $\lambda$ and $N$ are related as
\be
\rho {\lambda}^{-1} = \sqrt{\frac{N}{2} \left( \frac{N}{2} +1\right) } \ .
\ee

The states $|n_1 ,n_2 \rangle $ are eigenstates of the operator $X_3$,
whereas
$X_+$ and $X_-$ are rising and lowering operators respectively
\[
X_3 \ |n_1 ,n_2 \rangle \ = \ \frac{n_1 -n_2}{2} \ |n_1 ,n_2 \rangle \ ,
\]
\[
X_+ \ |n_1 ,n_2 \rangle \ = \ 2\sqrt{(n_1 +1) n_2} \ |n_1 +1,n_2 -1
\rangle \ ,
\]
\be
X_- \ |n_1 ,n_2 \rangle \ = \ 2\sqrt{n_1 (n_2 +1)} \ |n_1 -1,n_2 +1
\rangle \ .
\ee
Since $X_i : {\cal{F}}_N \rightarrow {\cal{F}}_N$, we have
\be
{\rm dim} {\cal A}_N \leq (N+1)^2 \ .
\ee

B) As a next step we extend the notions of integration and derivation to the
truncated case. The standard integral on $S^2$
\be
I_{\infty} (F) = \frac{1}{4\pi } \int d\Omega F(x) =
\frac{1}{4\pi } \int^{+\pi}_{-\pi}
d\varphi \int^{\pi}_0 \sin \theta d\theta F(\theta ,\varphi )
\ee
is uniquely defined if it is fixed for the monomials
$F(x)=x^l_+ x^m_- x^n_3$.
It is obvious that $I_{\infty} (x^l_+ x^m_- x^n_3 ) = 0$ for $l \not= m$, and
that $x^l_+ x^l_- x^n_3 =\rho^{2l+n} {\sin}^{2l} \theta {\cos}^n \theta$ is a
 polynomial
in $\cos \theta = x_3$. An easy calculation gives
\[
I_{\infty} (x_3^{2n+1} ) =0\ ,\ \ \ I_{\infty} (x_3^{2n} )
=\frac{\rho^{2n}}{2n+1} \ ,
\]
for $n=0,1,2,...\ $. Putting $\xi = \rho^{-1} x_3=\cos{\theta}$ we see that
\be
I_{\infty} ({\xi}^n) = \frac{1}{2} \int^{+1}_{-1} d\xi \ {\xi}^n \ .
\ee
These relations algebraically define the integration in ${\cal A}_{\infty}$.

In the non-commutative case we put
\be
I_N (F) = \frac{1}{N+1} {\rm Tr}[F(\hat x )]
\ee
for any polynomial $F(\hat x) \in \cal A_N$ in ${\hat x}_i , i=1,2,3$, where
the trace is taken in ${\cal F}_N$. Again, the integrals
$I({\hat x}^l_+ {\hat x}^m_- {\hat x}^n_3 ) = 0$ for $l \not= m$ since
\[
{\hat x}^l_+ {\hat x}^m_- {\hat x}^n_3 \ |n_1 ,n_2 \rangle \
\sim \ |n_1 +l-m,n_2 +m-l \rangle \ .
\]
Similarly as before, ${\hat x}^l_+ {\hat x}^l_- {\hat x}^n_3$ can be
expressed
using eqs. (5) and (6) as a polynomial in ${\hat x}_3$. The equation
\be
{\hat x}^n_3 \ |n_1 ,n_2 \rangle =
(\lambda \frac{n_1 -n_2}{2})^n \ |n_1 ,n_2 \rangle
\ee
gives
\be
I_N ({\hat x}^n_3 ) =
\sum_{k=0}^{N} \frac{\rho^n}{N+1} {\xi}_k^n \ ,
\ee
where ${\xi}_k = \sqrt{\frac{N}{N+2}} (\frac{2k}{N} -1)$. The formula (19)
can be rewritten as a Stieltjes integral with the stair-shape measure
$\mu (\xi )$ in the interval (-1,+1) with steps in the points ${\xi}_k$
\be
I_N ({\xi}^n ) = \int_{-1}^{+1} d\mu (\xi ){\xi}^n =\
\sum_{k=0}^{N} \frac{1}{N+1} {\xi}^n_k \ .
\ee
Obviously, $I_N ({\hat x}^{2n+1}_3 ) = 0$, and
\[
I_N ({\hat x}^{2n}_3 ) = \frac{\rho^{2n}}{(\frac{N}{2})^n (\frac{N}{2}+1)^n
 (N+1)}
\sum_{k=0}^{N} (\frac{2k-N}{2})^{2n} \ .
\]
Using the known formula (see e.g. [12], p. 597, eq. (16))
\[
\sum_{k=0}^{N} (k+a)^m = \frac{1}{m+1} [B_{m+1} (N+1+a) -B_{m+1} (a)] \ ,
\]
where $B_m (x)$ are Bernoulli polynomials, we obtain
\be
I_N ({\hat x}^{2n}_3 ) = \frac{\rho^{2n}}{2n+1} C(N,n)  \ .
\ee
Here,
\be
C(N,n) = \frac{B_{2n+1} (\frac{N}{2}+1) -B_{2n+1} (-\frac{N}{2})}{
(\frac{N}{2})^n (\frac{N}{2}+1)^n (N+1)}
\ee
represents a non-commutative correction. Since the Bernoulli polynomials are
normalized as
\[
B_m (x) = x^m \ + \ {\rm lower}\ {\rm powers} \ ,
\]
we see that
\be
C(N,n) = 1 + o(1/N) \ ,
\ee
i.e. in the limit $N \rightarrow \infty$ we recover the commutative result.

The scalar product in ${\cal A}_{\infty}$ can be introduced as
\be
(F_1 , F_2)_{\infty} = I_{\infty} (F_1^* F_2) \ ,
\ee
and similarly in ${\cal A}_N$ we put
\be
(F_1 , F_2)_N = I_N (F_1^* F_2) \ .
\ee

C) The vector fields describing motions on $S^2$ are linear combinations
(with the coeffitiens from ${\cal A}_{\infty}$) of the differential operators
acting on any $F \in {\cal A}_{\infty}$ as follows
\be
J_i F = \frac{1}{i} \ \varepsilon_{ijk} \ x_j \
\frac{\partial F}{\partial x_k}\ .
\ee
In particular,
\be
J_i x_j = i\ \varepsilon_{ijk} \ x_k \ .
\ee
The operators $J_i ,\ i=1,2,3$, satisfy in ${\cal A}_{\infty}$ the $su(2)$
algebra commutation relations
\be
[J_i, J_j ] = i \varepsilon_{ijk} J_k \ ,
\ee
or for $J_{\pm} = J_1 \pm iJ_2$ they take the form
\be
[J_3 ,J_{\pm} ]=\pm J_{\pm}\ ,\ \ [J_+ ,J_- ] = 2J_3 \ .
\ee
The operators $J_i$ are self-adjoint with respect to the scalar product (24).

In the non-commutative case  the operators $J_i$ act on any element $F$ from
the algebra ${\cal A}_N$ in the following way
\be
J_i F = [X_i \ ,F] \ .
\ee
In particular,
\be
J_i {\hat x}_j = i\ \varepsilon_{ijk} \ {\hat x}_k \ .
\ee
The operators $J_i$ satisfy $su(2)$ algebra commutation relations and are
self-adjoint with respect to the scalar product (25).

The functions
\be
\Psi_{ll} (\hat x ) = c_l \ {\hat x}^l_+ \ ,
\ee
are the highest  weight vectors in ${\cal A}_N$ for $l = 0,1,...,N$, since
\be
J_+ \Psi_{ll} (\hat x ) = {\lambda}^l \ [{\hat X}_+ , {\hat X}^l_+ ] = 0\ .
\ee
For all $l > N$ is ${\hat x}_+^l = 0$  in ${\cal A}_N$. The normalization
 factor
$c_l$ is fixed by the condition
\[
1 = \parallel \Psi_{ll} {\parallel}^2 = (\Psi_{ll} ,\Psi_{ll} )_N =
|c_l |^2 I_N({\hat x}^l_- {\hat x}^l_+ ) \ ,
\]
and is given by the formula ([12], p.618, eq. (36))
\be
\rho^{2l} c_l^2 = \frac{(2l+1)!!}{(2l)!!}
\frac{(N+1)N^l (N+2)^l (N-l)!}{(N+l+1)!} \ .
\ee
The second factor on the right hand side represents a non-commutative
correction. For $N \rightarrow \infty$ it approaches $1$. The other
normalized
functions $\Psi_{lm} , m = 0,\pm 1,...,\pm l$, in the irreducible
representation containing $\Psi_{ll}$ are given as
\be
\Psi_{lm} = \sqrt{\frac{(l+m)!}{(l-m)!(2l)!}}\ J^{l-m}_- \Psi_{ll} \ .
\ee
The normalization factor on the right hand side is the standard one
independent
of $N$. The functions $\Psi_{lm}$ are eigenfunctions of the operators $J^2_i$
and $J_3$:
\[
J^2_i \Psi_{lm} = l(l+1) \Psi_{lm} \ ,
\]
\be
J_3 \Psi_{lm} = m \Psi_{lm} \ .
\ee

We see that ${\cal A}_N$ contains all $SU(2)$ irreducible representations
with the
"orbital momentum" $l = 0,1,...,N$. The $l$-th representation has the
dimension
$2l+1$, and consequently
\be
{\rm dim}{\cal A}_N \geq \sum_{n=0}^N (2l+1) = (N+1)^2 \ .
\ee
Comparing this with eq. (14) we see that ${\cal A}_N$ contains no other
representations, i.e.
\be
{\cal A}_N = {\bigoplus}_{l=0}^N \ {\cal A}_{(l)} \ ,
\ee
where ${\cal A}_{(l)}$ denotes the representation space of the $l$-th
representation spanned by the functions $\Psi_{lm} , m = 0,\pm 1,...,\pm l$.
In particular, ${\rm dim}{\cal A}_N = (N+1)^2$.

\section{Scalar field on the  truncated sphere}
A) The Euclidean field action for a real self-interacting scalar field
$\Phi$ on a standard sphere $S^2$ is given as
\[
S[\Phi ] = \frac{1}{4\pi} \int_{S^2} d\Omega
[(J_i \Phi )^2 + \mu^2 (\Phi )^2 + V(\Phi )]
\]
\be
= I_{\infty} (\Phi J^2_i \Phi + \mu^2 (\Phi )^2 + V(\Phi )) \ ,
\ee
where
\be
V(\Phi ) = \sum_{k=0}^{2K} g_k \Phi^k \ ,
\ee
is a polynomial with $g_{2K} \geq 0$ (and we explicitely indicated the
mass term).

The quantum mean value of some polynomial field functional $F[\Phi ]$
is defined as the functional integral
\be
\langle F[\Phi ] \rangle =
\frac{\int D\Phi e^{-S[\Phi ]} F[\Phi ]}{\int D\Phi e^{-S[\Phi ]}} \ ,
\ee
where $D\Phi = \prod_x d\Phi (x)$. Alternatively, we can expand the field
into
spherical functions
\be
\Phi (x) = \sum_{l=0}^{\infty} \sum_{m=-l}^{+l} a_{lm} Y_{lm} (x)
\ee
satisfying
\[
J^2_i Y_{lm} \ = \ l(l+1) Y_{lm} \ .
\]
Here the complex coefficients $a_{lm}$ obey
\be
a_{l,-m} = (-1)^m a^*_{lm} \ ,
\ee
what guarantees the reality condition
$\Phi^* ({\hat x})$ $=$ $\Phi ({\hat x})$. We can
put $D\Phi$ $=$ $\prod_l da_{l0}\prod_{lm} da_{lm} da^*_{lm} \ $,
$\ l=0,1,...,N$,$\ m = 1,...,l$. Both
expressions for $D\Phi$ are only formal. The measure in the functional
integral can be mathematically rigorously defined (see e.g. [13]) but we
shall
not follow this direction.

Such problems do not appear in the non-commutative case, where the
scalar field $\Phi ({\hat x})$ is  an element of the algebra ${\cal A}_N$,
and consequently it can be expanded as
\be
\Phi ({\hat x}) = \sum_{l=0}^{N} \sum_{m=-l}^{+l}
a_{lm} \Psi_{lm} ({\hat x}) \ ,
\ee
where $\Psi_{lm} ({\hat x})$ satisfy in ${\cal A}_N$ the equation
\[
J^2_i \Psi_{lm} \ = \ l(l+1) \Psi_{lm} \ ,
\]
and are orthonormal with respect to the scalar product (25). The coefficients
$a_{lm}$ are again restricted by the condition (43).

The action in the non-commutative case is defined as
\be
S[\Phi ] =
 I_N (\Phi J^2_i \Phi + \mu^2 (\Phi )^2 + V(\Phi )) \ ,
\ee
and it is a polynomial in the variables
$a_{lm} , l=0,1,...,N,\ m = 0,\pm 1,...,\pm l$. The measure
$D\Phi = \prod_l da_{l0}\prod_{lm} da_{lm} da^*_{lm} \ ,\ l=0,1,...,N,\ m =
 1,..., l$, in the
quantum mean value (41) is the usual Lebesgue measure, since now the product
is finite. The quantum mean values are well defined for any analytic
functional $F[\Phi ]$.

Under rotations
\be
{\hat x}_i \rightarrow {\hat x}'_i =
\sum_j R_{ij} (\alpha ,\beta ,\gamma ) {\hat x}_j
\ee
specified by the Euler angles $\alpha ,\beta ,\gamma$, the field
transforms as
\be
\Phi ({\hat x}) \rightarrow \Phi ({\hat x}') =
\sum_{l=0}^{N} \sum_{m=-l}^{+l} a_{lm} \Psi_{lm} ({\hat x}') \ .
\ee
Using the transformation rule for the functions $\Psi_{lm}$ (see e.g. [15])
\be
\Psi_{lm'} ({\hat x}') = \sum_{m'} D^l_{m'm} (\alpha ,\beta ,\gamma )
\Psi_{lm} ({\hat x}) \ ,
\ee
we obtain the transformation rule for the coefficients $a_{lm}$
\be
a_{lm} \rightarrow a'_{lm'} =
\sum_{m} D^l_{m'm} (\alpha ,\beta ,\gamma ) a_{lm} \ .
\ee
The last equation is an orthogonal transformation not changing the measure
$D\Phi $ (see e.g. [14]).

The Schwinger functions we define as follows
\be
S_n (F) = \langle F_n [\Phi ] \rangle \ ,
\ee
where
\be
F_n[\Phi]=\sum \alpha_{l_1 m_1\dots l_n m_n} a_{l_1 m_1}\dots a_{l_n
m_n}\equiv
\sum\alpha_{l_1 m_1\dots l_n m_n}(\Psi_{l_1
m_1},\Phi )_N\dots (\Psi_{l_n m_n},\Phi )_N \ .
\ee
The functions (49) satisfy the following  Osterwalder-Schrader axioms:

(OS1) {\it Hermiticity}
\be
S^*_n (F) = S_n (\Theta F) \ ,
\ee
where $\Theta F$ is an involution
\[
\Theta F_n [\Phi] = \sum\alpha^*_{l_1 -m_1\dots l_n
-m_n}(-1)^{m_1+\dots m_n} a_{l_1 m_1}\dots a_{l_n m_n}\ .
\]

(OS2) {\it Covariance}
\be
S_n (F) = S_n ({\cal R}F) \ ,
\ee
where ${\cal R}F$ is a mapping induced by Eq. (49).

(OS3) {\it Reflection positivity}
\be
\sum_{n,m\in {\cal I}} S_{n+m} (\Theta F_n \otimes F_m ) \geq 0 \ .
\ee

(OS4) {\it Symmetry}
\be
S_n (F) = S_n (\pi F) \ ,
\ee
where $\pi F$ is a functional obtained from $F$ by arbitrary permutation
of $a_{lm}$'s in Eq. (51).

{\it Note}:
The positivity axiom (53) can be rewritten as $\langle F^* F
\rangle \geq 0$, $F=\sum_{n\in{\cal I}}F_n$. In fact, the standard
formulation of (OS3) axiom requires the specification of the
support of the functionals $F_n$. In our case the axiom holds in the
"strong" sense , i.e. without the specification. We expect, however,
that in the continuum limit ($N\to\infty$) the issue will emerge.
We do not include the last Osterwalder-Schrader axiom - the cluster property,
since the compact manifold requires a special treatment (however, it can be
recovered in the limit when the radius of the sphere grows to infinity, but
this goes beyond the presented scheme).

B) In many practical applications the perturbative results are sufficient.
Interpreting the term $V(\Phi )$ as a perturbation, we present below
as an illustration the Feynman rules for the model in question. We give the
Feynman rules in the $(lm)$-representation defined by the expansios (42) and
(43). The diagrams are constructed from

(i) {\it External vertices}
assigned to any operator $a_{lm}$ appearing in the functional $F[\Phi ]$.

(ii) {\it Internal vertices}
given by the expansion of $V(\Phi)$ in terms of
$a_{l_1 m_1}...a_{l_k m_k}$.

This gives the following Feynman rules:

(a) {\it Propagator}
\be
2\langle a_{lm}a^*_{l'm'}\rangle= \ \frac{1}{l(l+1)+\mu^2}
\delta_{l'l} \delta_{m'm} \ ,
\ee
where the admissible values of $l$ and $m$ for ${\cal A}_{\infty}$ are
$l=0,1,2,...,\ m = 0,1,...,l$, whereas in the case of ${\cal A}_N$
they are $l=0,1,...,N,\ m = 0,1,...,l$.

(b) {\it Vertex}

\be
V_{l_1 m_1 ,...,l_k m_k } \ = \ g_k I_{\infty} (Y_{l_1 m_1}...Y_{l_k m_k } )
\ {\rm for} \ {\cal A}_{\infty}  \ ,
\ee
\be
V_{l_1 m_1 ,...,l_k m_k } \ = \ g_k I_N (Y_{l_1 m_1}...Y_{l_k m_k } )
\ {\rm for} \ {\cal A}_N  \ ,
\ee

(c) Finally the summation over all {\it internal} indices should be
performed.

This procedure leads for ${\cal A}_{\infty}$ finite Feynman diagrams except
the diagrams containing the tadpole contribution
\[
T_{\infty} \ \equiv \sum_{lm}\langle a_{lm}a^*_{lm}\rangle \sim
 \sum_{l=0}^{\infty}\sum_{m=-l}^{l} \frac{1}{l(l+1)+\mu^2} \ =\ \infty \ .
\]
This divergence is closely related to the divergence of the propagator
\[
G(x,y)\ =\ \sum_{lm} \frac{1}{l(l+1)+\mu^2} Y_{lm} (x) Y^*_{lm} (y)
\]
in the x-representation at points $x=y$. This requires, of course, the
regularization of $G(x,y)$, which is, in our case, simply
 a cut-off in the $l$-summations. Indeed,
for ${\cal A}_N$ all diagrams are obviously finite (since all summations
are finite). In particular the tadpole contribution reads
\[
T_N \ =\  \sum_{l=0}^{N}\sum_{m=-l}^l \frac{1}{l(l+1)+\mu^2} \sim \ \ln N \ .
\]

For practical applications an effective method for the calculation of
vertex coefficients $V_{l_1 m_1 ,...,l_k m_k }$ is
needed, both in the standard and non-commutative cases. We shall describe
the latter one. Since the multiplication by $\Psi_{lm}$ acts in the algebra
${\cal A}_N$ as an
 irreducible tensor operator, we can apply the Wigner-Eckart
theorem. Then the product $\Psi_{l_1 m_1} (\hat x ) \Psi_{l_2 m_2} (\hat x )$
can be expressed as
\be
\Psi_{l_1 m_1} (\hat x ) \Psi_{l_2 m_2} (\hat x ) =
\sum_{l=|l_1 -l_2 |}^{l_1 +l_2}
(l_1 m_1 ,l_2 m_2 |l m)\ (l_1 l_2 \parallel l)\ \Psi_{lm} (\hat x ) \ ,
\ee
where $m = - m_1 + m_2$, $\ (l_1 m_1 ,l_2 m_2 |l m)$ is a Clebsch-Gordon
coefficient, and the symbol $(l_1 l_2 \parallel l)$ denotes the so called
reduced matrix element (and depends on the particular algebra in question).
Introducing the non-commutative Legendre polynomials
$P_l (\xi ) = \Psi_{l0} (\hat x ),\ \xi = \rho^{-1}{\hat x}_3$, the previous
equation
leads to the coupling rule
\be
P_{l_1} (\xi ) P_{l_2} (\xi ) =
\sum_{l=|l_1 -l_2 |}^{l_1 +l_2}
(l_1 0,l_2 0 |l 0)\ (l_1 l_2 \parallel l)\ P_l (\xi ) \ .
\ee
The repeated application of (59) then allows to calculate the required
vertices.

{\it Note:} The well known explicit formula for the usual Legendre
polynomials allows us to calculate the reduced matrix elements
\[
(l_1 l_2 \parallel l) = (l_1 0,l_2 0 |l 0)
\]
enterring the coupling rule in the algebra ${\cal A}_{\infty}$ in terms of a
particular Clebsch-Gordon coefficients. Similarly, the explicit formula
for the non-commutative Legendre polynomials presented in the Appendix
allows to deduce the reduced matrix elements enterring the coupling rule
in the algebra ${\cal A}_N$.

\section{Concluding Remarks}
We have  demonstrated above that the interacting
scalar field on the non-commutative sphere
represents a quantum system which has the following properties:

1) The model has a full space symmetry - the full symmetry under
isometries (rotations) of the sphere $S^2$. This is exactly the same
symmetry as the interacting scalar field on the standard sphere has.

2) The field has only a finite number of modes. Then the number of degrees
of freedom is finite and this leads to the non-perturbative
UV-regularization, i.e. all quantum mean values of polynomial field
functionals are well defined and finite.

Consequently, all Feynman diagrams in the perturbative expansion are finite,
even the diagrams containing the tadpole diagram which are divergent in the
model on a standard sphere. Technically, the tadpole is finite due to the
cut-off in the number of modes. In our approach the
UV cut-off in the number of modes is supplemented with a highly non-trivial
vertex modification (compare eqs. (57) and (58)). Moreover, our
UV-regularization is non-perturbative and is completely determined by the
algebra ${\cal A}_N$. It is originated by the short-distance structure of
the space, and does not depend on the field action of the model in question.
{}From the presented point of view, it would be desirable
to analyze a quantization of the models on a non-commutative sphere $S^2$
containing spinor, or gauge fields. In the standard case such models  have
a more complicated structure of divergencies. It is evident,
that our approach will lead again to a non-perturbative UV-regularization.
The usual divergencies will appear only in the limit $N \rightarrow \infty$.
It would be very interesting to isolate the large $N$ behaviour
non-perturbatively. By this we mean the Wilson-like approach
in which the renormalization group flow in the space of Lagrangeans
is studied. This can lead to the better
understanding of the origin and properties of divergencies in the quantum
field theory. Another interesting direction would consist in making
connection with the matrix models where, from the technical point of
view, very similar integral have been studied.
 We strongly believe, that qualitatively just the same situation
will repeat on the four-dimensional sphere $S^4$ too. Investigations in all
these directions are under current study.

{\bf Ackonwledgement}\newline
We are grateful to A. Alekseev, L. \'{A}lvarez-Gaum\'{e}, M. Bauer,
A. Connes,
V. \v{C}ern\'{y}, T. Damour,
 J. Fr\"{o}hlich, J. Ft\'{a}\v{c}nik, K. Gaw\c edzki,
J. Hoppe, B. Jur\v{c}o, E. Kiritsis, C. Kounnas, M. Rieffel, R. Stora
and D. Sullivan for useful discussions.
The part of the research of C.K. has been done at I.H.E.S. at
Bures-sur-Yvette and of C.K. and P.P. at the Schr\"odinger Institute
in Vienna. We thank both of them for hospitality.
\\

{\bf Appendix}\\
The truncated Legendre polynomials
\[
P_l(\xi ) = \xi^l a^l_0 + \xi^{l-2} a^l_1 + ...\ ,\ \ l=0,1,...,N\ ,
\]
we define as orthonormal polynomials with respect to the scalar product
\[
(P_l ,P_m )_N = I_N (P_l P_m ) = \delta_{lm} \ .
\]
Here the non-commutative integral is given as (see eq. (19))
\[
I_N ({\xi}^n ) =
\sum_{k=0}^{N} \frac{1}{N+1} {\xi}_k^n \ ,
\]
where ${\xi}_k = \sqrt{\frac{N}{N+2}} (\frac{2k}{N} -1)$. The polynomials
$P_l(\xi )$ can be obtained from the reccurence relation
\[
P_{m+1} (\xi ) = \frac{1}{a_m} [\xi P_m (\xi ) -c_m P_{m-1} (\xi )]\ ,
\]
where $c_m = I(\xi P_m P_{m-1} )$ and $a_m = \sqrt{I_N(\xi^2 P^2_m )-c^2_m}$.

The truncated spherical functions $\Psi_{lm} (\hat x )$ satisfy in
${\cal A}_N$
equation
\[
J^2_i \Psi_{lm} (\hat x ) = l(l+1) \Psi_{lm} (\hat x ) \ .
\]
Putting $P_l(\xi ) = \Psi_{l0} (\hat x )\ , \xi = {\hat x}_3$, the last
equation reduces to a difference equation for the truncated Legendre
polynomials
\[
(1-\xi^2 )
\frac{P_l(\xi +\lambda )-2P_l(\xi )+P_l(\xi -\lambda )}{\lambda^2}\
\]
\[
+\ 2\xi \frac{P_l(\xi +\lambda )-P_l(\xi -\lambda )}{2\lambda}\ +\ l(l+1)
P_l(\xi ) = 0 \ ,
\]
where $\lambda =2/ \sqrt{N(N+2)}$. This equation leads to the reccurence
relation for the coefficients $a^l_s$ appearing in the Legendre polynomials:
\[
a^l_s = -\frac{1}{s(2l-2s+1} \sum_{r=0}^{s-1} a^l_r
[({}^{l-2r}_{l-2s} )\ -{\lambda}^2 \ ({}^{l-2r+1}_{l-2s+1} )]
\ \lambda^{2s-2r-2} \ .
\]
In the limit $N \rightarrow \infty$ (or equivalently $\lambda \rightarrow 0$)
all formulas reduce to the standard expressions valid for usual Legendre
polynomials.


\newpage
\begin{thebibliography}{99}
\bibitem{1} A. Connes, Publ. IHES {\em 62} (1986) 257.
\bibitem{2} A. Connes, Geometrie Noncommutative (Inter Editions, Paris
1990).
\bibitem{3} M. Dubois-Violette, C. R. Acad. Sci. Paris {\em 307}, Ser. I
(1988) 403.
\bibitem{4} M. Dubois-Violette, R. Kerner and J. Madore, Journ. Math. Phys.
{\em 31} (1990) 316.
\bibitem{5} R. Coquereaux, G. Esposito-Farese and G. Vailant, Nucl. Phys.
{\em B353} (1991) 689.
\bibitem{6} A.H. Chamsedine, G. Felder and J. Fr\"{o}hlich,
{\it Gravity in the non-commutative geometry}, preprint
ETH-TH/1992-18.
\bibitem{7} J. Madore, Journ. Math. Phys. {\em 32} (1991) 332.
\bibitem{8} H. Grosse and P. Pre\v{s}najder, Lett. Math. Phys.
{\em 28} (1993) 239.
\bibitem{9} H. Grosse and J. Madore, Phys. Lett. {\em B283} (1992) 218.
\bibitem{10} H. Grosse, C. Klim\v{c}\'{\i}k  and P. Pre\v{s}najder,
{\it Field theory on truncated superspace} - to be published.
\bibitem{11} H. Grosse, C. Klim\v{c}\'{\i}k  and P. Pre\v{s}najder,
{\it Finite gauge model on truncated sphere} - to be published.
\bibitem{12} A. P. Prudnikov, Yu. A. Brytshkov and O. I. Maritshev,
{\it Integrals and Series}, Publ. by Nauka, Moscow 1981 (in Russian).
\bibitem{13} B. Simon, {\it The $P(\Phi )_2$ Euclidean (Quantum) Field
Theory}, Princeton University Press, Princeton - New Jersey 1974.
\bibitem{14} N. Ya. Vilenkin, {\it Special Functions and the Theory of
Group Representations}, Publ. by Nauka, Moscow 1965 (in Russian).
\end{thebibliography}
\end{document}